\documentclass[fleqn,usenatbib]{mnras}

\usepackage{newtxtext,newtxmath}

\usepackage{xcolor}
\usepackage{url}
\usepackage{pdflscape}
\usepackage{soul}
\usepackage[pagewise]{lineno}
\usepackage[T1]{fontenc}

\DeclareRobustCommand{\VAN}[3]{#2}
\let\VANthebibliography\thebibliography
\def\thebibliography{\DeclareRobustCommand{\VAN}[3]{##3}\VANthebibliography}

\usepackage{graphicx}	
\usepackage{amsmath}	
\usepackage{subfigure}

\title[Neutrino emission from very massive stars]{The neutrino emission from thermal processes in very massive stars in the local universe}

\author[N, Yusof]{
N, Yusof$^{1}$ \thanks{E-mail: norhaslizay@um.edu.my},
H. A. Kassim$^{1}$,
L.G. Garba$^{1,2}$ and
N.S. Ahmad $^{1}$ 
\\
$^{1}$Department of Physics, Faculty of Science, University of Malaya, 50603 Kuala Lumpur, Malaysia\\
$^{2}$Department of Physics, Faculty of Science, Yusuf Maitama Sule
University, Kano 3220, Nigeria
}

\date{Accepted XXX. Received YYY; in original form ZZZ}

\pubyear{2020}

\begin{document}
\label{firstpage}
\pagerange{\pageref{firstpage}--\pageref{lastpage}}
\maketitle

\begin{abstract} We present a new overview of the life of very massive stars (VMS) in terms of  neutrino emission from thermal processes: pair annihilation, plasmon decay, photoneutrino process, bremsstrahlung and recombination processes in burning stages of selected VMS models. We use the realistic conditions of temperature, density, electron fraction and nuclear isotropic composition of the VMS. Results are presented for a set of progenitor stars with mass of 150, 200 and 300 M$_\odot$ Z=0.002 and 500 M$_\odot$ Z=0.006 rotating models which are expected to explode as a pair instability supernova at the end of their life except the 300 M$_\odot$ would end up as a black hole. It is found that for VMS, thermal neutrino emission occurs as early as towards the end of hydrogen burning stage due to the high initial temperature and density of these VMS. We calculate the total neutrino emissivity, $Q_\nu$ and luminosity, $L_\nu$ using the structure profile of each burning stages of the models and observed the contribution of photoneutrino at early burning stages (H and He) and pair annihilation at the advanced stages. Pair annihilation and photoneutrino processes are the most dominant neutrino energy loss mechanisms throughout the evolutionary track of the VMS. At the O-burning stage, the neutrino luminosity $\sim 10^{47-48}$ erg/s depending on their initial mass and metallicity are slightly higher than the neutrino luminosity from massive stars. This could shed light on the possibility of using detection of neutrinos to locate the candidates for pair instability supernova in our local universe. 
\end{abstract}




\section{Introduction}
Neutrinos play an important role in the massive stars evolution through neutrino energy loss or neutrino cooling. This process is essential in keeping the stars burning later in their life in order to prevent early gravitational collapse in its core. For massive stars, the neutrino processes are governed by four main processes; pair production, plasmon decay, photoneutrino emission and bremsstrahlung  and they have been studied extensively for decades.  \citep{beaudet1967energy,dicus1976neutrino,braaten1993neutrino,munakata1985neutrino,itoh1989neutrino,itoh1992neutrino,1996ApJS..102..411I,esposito2003neutrino,guo2016spectra}. \cite{1996ApJS..102..411I} have included a treatment on the recombination process for nonrelativistic electrons which would be important at low temperature-density regimes. During the evolution of massive (or very massive) stars, the thermal processes taking place more pronounced during the start of carbon burning phase and it will continue until their Fe core collapses and proceeds to the supernova explosion. For each advanced burning stages, the lifetime of the burning stages is shorter and this means more neutrinos are produced and released to the interstellar medium. 

For normal massive stars, the advanced burning stages are listed as follows: carbon, neon, oxygen and finally silicon burning before their core  collapses due to the gravitational contraction \citep{heger2003massive}. Pre-supernova neutrinos produced from thermal processes of massive stars and the possibility of pre-supernova neutrino detection were first discussed by \cite{odrzywolek2004detection, kutschera2009presupernovae} by detecting signals from the pair annihilation neutrinos emission of the 20 M$_\odot$. Pre-supernova neutrino emission from ONe cores using a realistic stellar evolution model has been discussed by \cite{kato2015pre} and further discussion using all flavors of the neutrinos and possibility of detection can be found for the electron capture supernova progenitors \citep{kato2017neutrino}. \cite{patton2017presupernova,patton2017neutrinos} gave a discussion on the calculation of neutrino emissivities that includes contributions from $\beta$ decay and thermal processes for a set of progenitor stars between 15 M$_\odot$ and 30 M$_\odot$. \cite{yoshida2016presupernova} shows that the pre-supernova neutrinos can be used to investigate the neutrino events from neutrino observatories. \cite{patton2017presupernova} do not include bremsstrahlung neutrinos. In massive stars, the temperature and density regimes are not favourable for bremsstrahlung process.

But for very massive stars, stars with initial mass greater than 100 M$_\odot$, the burning process is shorten until oxygen burning phase. This is due to the production of the  electron-positron ($e{^-}e{^+}$) pairs and their annihilation that destroys the structure of VMS to produce a pair instability supernova (PISN)\citep{barkat1967dynamics,rakavy1967instabilities, fryer2001pair, kasen2011pair}. The existence of these supernovae is important to both chemical evolution \citep{karlsson2008uncovering} and galaxy formation  perspectives \citep{whalen2008destruction}. However, the existence of 
PISN has not yet been found in any metal-poor stars until this date but for the local universe, some possible candidates have been identified 
\cite{gal2009supernova,cooke2012superluminous}. Since in the evolution of VMS, the end product has the possibility of having very high luminosity and temperature, it is important to understand how much neutrinos can be released during the pre-supernova stage since neutrinos can be an important messenger of supernova. \cite{wright2017neutrino} demonstrated that the possible neutrino signal from pair-instability supernova for progenitor with Z=10$^{-3}$  can be used to design future detectors. Besides pair instability, very massive stars also can be progenitors of pulsation pair instability supernovae (PPISNe) and black holes, and some recent work on neutrino signal from \cite{leung2020pulsational} study the neutrino emission and the expected detection count for different terrestrial 
detectors. 

Stars with mass greater than 100 M$_\odot$ have been observed within the Large Margellanic Cloud (see \cite{crowther2010r136}) and it is possible to end their life as PISN and located at distance of $d=50$ kpc \citep{pietrzynski2013eclipsing}. The new-generation detector, Hyper-Kamiokande  is envisaged to be able to detect the PISN event at the LMC location since it was reported that Hyper-Kamiokande is able to detect events as far as 100 kpc \citep{migenda2020supernova, abe2021supernova}.

In this paper, we explore the neutrino production in very massive stars in particular very massive stars in the local universe at Small Margellanic Cloud (SMC) and Large Margellanic Cloud (LMC) metallicities.  This can be the signature of the production of the pair electron-positron ($e{^-}e{^+}$) annihilation that contributes to the explosion of pair instability supernova. We also focus on the neutrino thermal energy distribution and estimate how much energy from the neutrino energy loss will supersede the energy from the nuclear reaction rates.  We here extend the discussion from \cite{yusof2013evolution} for the neutrino energy loss with selected VMS models from the same reference. We explore in this paper on how do neutrinos from very massive stars could enrich or contribute to the production of pair instability supernovae and black holes by a given mass. 

\section{Neutrino Production in the Evolutionary Phase}

Massive stars (> 8M$_\odot$), undergo the production of heavy elements beyond helium burning due to the increase of temperature and density during their evolutionary phases. When carbon fusion starts, the temperature and density would further increase and the fusion process is then followed by Ne, O and Si fusions at the core of the stars. Each burning phases are faster than the previous one where O-burning takes a few months while Si core burning lasts only a few days \citep{woosley2002evolution}. However, for very massive stars (>100 M$_\odot$), their lifetimes are shorter than their lighter siblings wherein their burning stages most probably would end at the O-burning stage \citep{heger2003massive} due to the impact of the $e^+e^-$ pair production during the O-burning. This however will depend on few other factors such as mass loss and metallicity \citep{yusof2013evolution}.

To describe very massive stars environment, we revisited the thermal neutrino processes and analyse how do these processes will impact on the very massive stars evolution since they have different range in temperature and density profiles compared to the massive stars. Following \cite{guo2016spectra}, we present the regions where any of the five thermal neutrino processes dominates by at least 90\% of the total neutrino loss. Included are the evolutionary tracks of the selected VMS models.

\begin{figure}
	\includegraphics[width=\columnwidth]{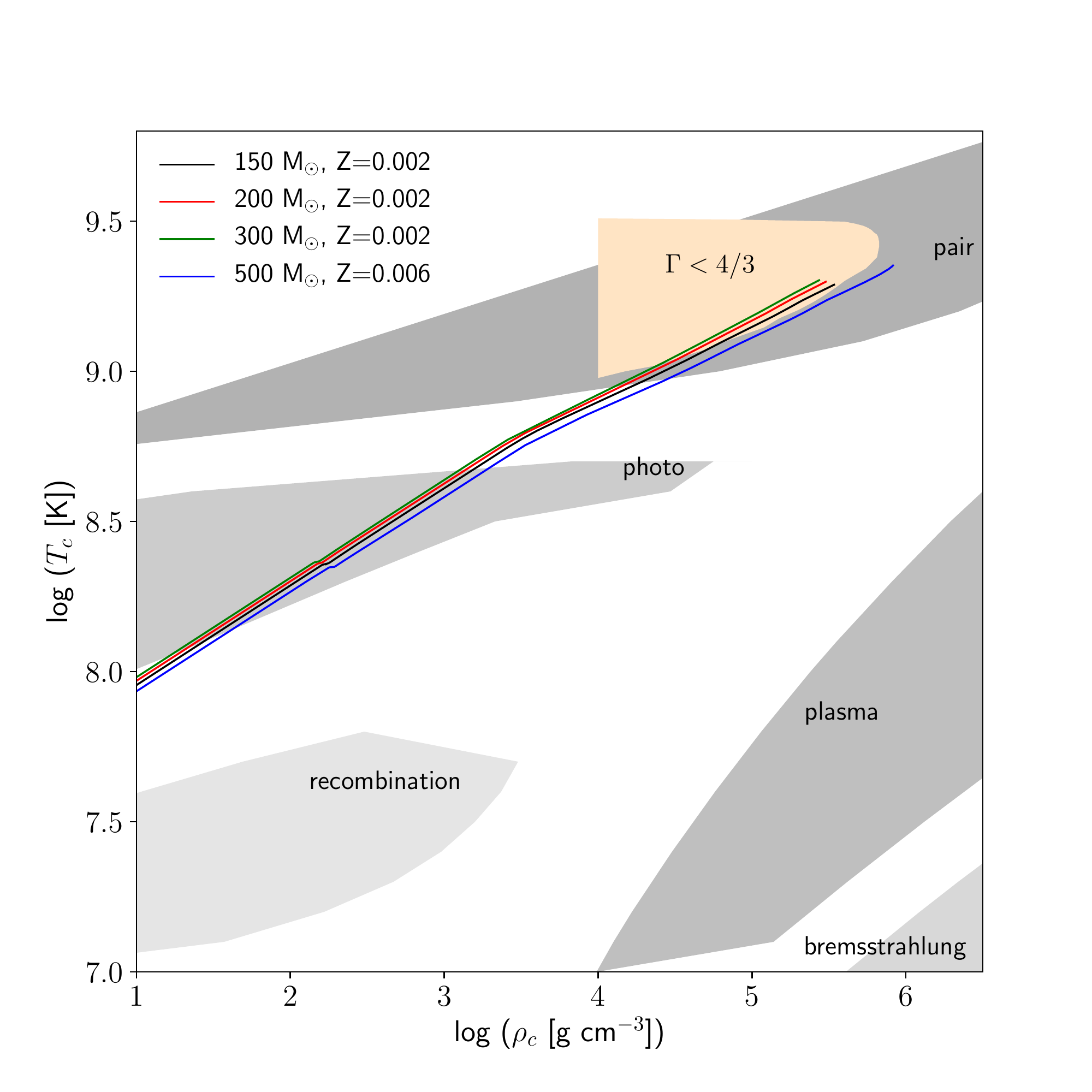}
    \caption{The evolutionary tracks of central temperature and density from ZAMS until at least O-burning stage of all VMS models. Also shown the neutrino processes in their temperature-density dominance region boundaries. The light orange region indicates the instability region, $\Gamma<4/3$.}
    \label{fig:neuprocess_general}
\end{figure}

\subsection{Thermal Processes}
The important neutrino processes in stellar environment discussed in the literature are pair annihilation, photoneutrino production, plasmon decay, bremsstrahlung and recombination processes. \cite{1996ApJS..102..411I} and in the series of papers produced by the same group have provided the details of the total emission for a broad range of temperatures and densities based on the well known Weinberg-Salam theory in the Standard Model of particle physics. These include neutrino emissivity that can be read from tables and approximation (fitting) equations that can be utilized directly in stellar models or apply in post-processing work. The total neutrino energy loss or neutrino emissivity, $Q_\nu$ is the sum all of the losses from the five different neutrino processes:
\begin{equation}
    Q_\nu =  Q_\mathrm{pair} + Q_\mathrm{photo} + Q_\mathrm{plasma}+Q_\mathrm{brems}+Q_\mathrm{recomb}
\end{equation}
where $ Q_\mathrm{pair}, Q_\mathrm{photo}, Q_\mathrm{plasma}, Q_\mathrm{brems} $ and $Q_\mathrm{recomb}$ are the neutrino emissivities from the pair annihilation, photoneutrino production, plasmon decay, bremsstrahlung and recombination processes respectively. In the following sections, we summarize the formalism found in \cite{1996ApJS..102..411I} that is relevant in our calculations.

\subsubsection{Pair annihilation}

Neutrino energy loss rate from the pair neutrino process, ($e^+ e^- \rightarrow{ v \bar{v}}$) is expressed as \citep{munakata1985neutrino,itoh1989neutrino}:

\begin{multline}
     Q_\mathrm{pair}=\frac{1}{2}[C{_V^2}+C{_A^2}+n(C'{_V^2}+C'{_A^2})]Q^+_{\mathrm{pair}}+\\
     \frac{1}{2}[C{_V^2}-C{_A^2}+n(C'{_V^2}-C'{_A^2})]Q^-_{\mathrm{pair}}
\end{multline}

\begin{equation}
    C_V=\frac{1}{2}+2 \sin^2\theta_W, C_A=\frac{1}{2}
\end{equation}
\begin{equation}
    C'_V=1-C_V, C'_A=1-C_A
\end{equation}
where $\theta_W$ is the Weinberg angle or better known as the weak mixing angle in the quark sector and $n$ is the number of flavors of the neutrinos. In all of our calculations, we have included all the experimentally observed three neutrino flavors. Suffice to note that, effects of the flavor-changing neutrino oscillations are not included in the present work.

Pair annihilation process is dominant at density-temperature region $10^0 \leq \rho/\mu_e (\mathrm{gcm})^{-3} \leq 10^{14}$, $10^7 \leq T$(K) $\leq 10^{11}$. This is well within our range of $T-\rho$ region occupied by our four VMS models as depicted by Figure \ref{fig:neuprocess_general}. Pair-neutrino process dominates over all other thermal neutrino processes at central temperatures $10^{8.8}$K $\lesssim  T_c \lesssim  10^{9.3}$K and central densities $10^{4}\mathrm{gcm}^{-3} \lesssim  \rho_c \lesssim  10^{6} \mathrm{gcm}^{-3}$ starting from carbon burning and beyond (see Table \ref{tab:schematic}). Our models do not evolved beyond the upper limit of these ranges as the evolution of these VMS models stop at core O-burning. As shown in \cite{1996ApJS..102..411I}, plasma neutrino process would grow in importance and pair neutrino contribution starts to decline at higher temperatures and densities but our models do not reach these extreme regions thus avoiding large errors that could dominate over results that are inherent in the approximation equations. As pair neutrino process is the most important contributor towards the energy loss in the VMS, it is important to keep the errors to the minimum possible.

\subsubsection{Photoneutrino process}

For photoneutrino process, ($\gamma e \rightarrow{e v \bar{v}}$) the energy loss is represented by:

\begin{multline}
    Q_\mathrm{photo}=\frac{1}{2}[(C{_V^2}+C{_A^2})+n(C'{^2_V}+C'{^2_A})]Q^+_\mathrm{photo}\\
    -\frac{1}{2}[(C{_V^2}-C{_A^2})+n(C'{^2_V}-C'{^2_A})]Q^-_\mathrm{photo}
\end{multline}

This process is expected to dominate at temperature-density region $10^0 \leq \rho/\mu_e (\mathrm{gcm})^{-3} \leq 10^{11}$, $10^7 \leq T$(K) $\leq 10^{11}$. As we can observe in Figure \ref{fig:neuprocess_general}, the photoneutrino process is dominant in the VMS models at lower central temperature region $10^{8}$K $\lesssim  T_c \lesssim  10^{8.5}$K and central density $\rho_c \lesssim  10^{4.5} \mathrm{gcm}^{-3}$ than the pair process. The approximation emissivity equation is fitted accurately in these regions \citep{1996ApJS..102..411I}. Photoneutrino process dominates prior to carbon burning stage.  

\subsubsection{Plasmon decay}

The neutrino energy loss rates in plasmon decay or plasma neutrino process, ($\gamma \rightarrow v \bar{v}$) is  written as

\begin{equation}
    Q_\mathrm{plasma}=(C{_V^2}+n C'{_V^2})Q_V
\end{equation}

and
\begin{equation}
    Q_V=Q_L+Q_T
\end{equation}
where $Q_L$ and $Q_T$ are the contributions of the longitudinal plasmon and transerve plasmon \citep{1996ApJS..102..411I} respectively. Plasma neutrino loss is important at very much higher temperature-density range than encountered by the VMS models as shown by Figure \ref{fig:neuprocess_general}. It only starts to be sensitive to the increase in temperature at $T > 10^8$ K and will not dominate over the pair neutrino in the evolutionary track of the VMS models.

\subsubsection{Bremsstrahlung process}
The energy rate of the bremsstrahlung process, ($e^- (Ze) \rightarrow (Ze)e^- v \bar{v}$)  has been calculated by \cite{itoh1983neutrino,munakata1987neutrino} and the earlier work by  \cite{dicus1976neutrino}. Bremsstrahlung process contributes the least amongst the five thermal neutrino processes as seen in Figure \ref{fig:neuprocess_general} consistent with  the literature for massive stars \citep{odrzywolek2004detection, patton2017presupernova} due to the temperature being too low. Nevertheless, we include this process and analyse the impact in the very massive stars evolution. 

There are five interactions in bremsstrahlung process and they are weakly degenerate electrons, liquid metal phase, binary ion mixture, low temperature quantum corrections in liquid metal phase and crystalline lattice phase. Since our models are at pre-supernova stage, it would be safe to assume the weakly degenerate electron is the most important in the star. For weakly degenerate electrons the energy loss is given as:

\begin{multline}
    Q_\mathrm{gas}=0.5738(\mathrm{erg} \mathrm{cm}^{-3}\mathrm{s}^{-1}(Z^2/A)T^6_8 \rho \\
    \times {\frac{1}{2}[(C^2_V+C^2_A)+n(C'^2_V+C'^2_A)]F_\mathrm{gas}} \\
     \times {\frac{1}{2}[(C^2_V-C^2_A)+n(C'^2_V-C'^2_A)]G_\mathrm{gas}}
\end{multline}
where the complete definition of $F_\mathrm{gas}$ and $G_\mathrm{gas}$ can be obtained from \cite{1996ApJS..102..411I} and here $Z$ is the atomic number.

\subsubsection{Recombination process}

The formula for this process is as given by \cite{1996ApJS..102..411I} and is expressed as:
\begin{multline}
    Q_\mathrm{recomb}=[(C_V^2+\frac{3}{2}C^2_A)+n(C'^2_V+\frac{3}{2}C'^2_A)] \\
    \times 2.649 \times 10^{-18} \frac{Z^{14}}{A} \rho \frac{1}{e^{\zeta+\nu}+1}J (\mathrm{erg}/ \mathrm{cm}^{3}\mathrm{s})
\end{multline}
where $Z$ is the atomic number. For recombination process, this calculation is valid for non-relativistic electrons and satisfies the limits on the density  $\rho/ \mu_e \leq 10^6 \mathrm{g}\mathrm{cm}^{-3}$  and temperature, $T \leq 6 \times 10^9 \mathrm{K}$.

\begin{figure}
	\includegraphics[width=\columnwidth]{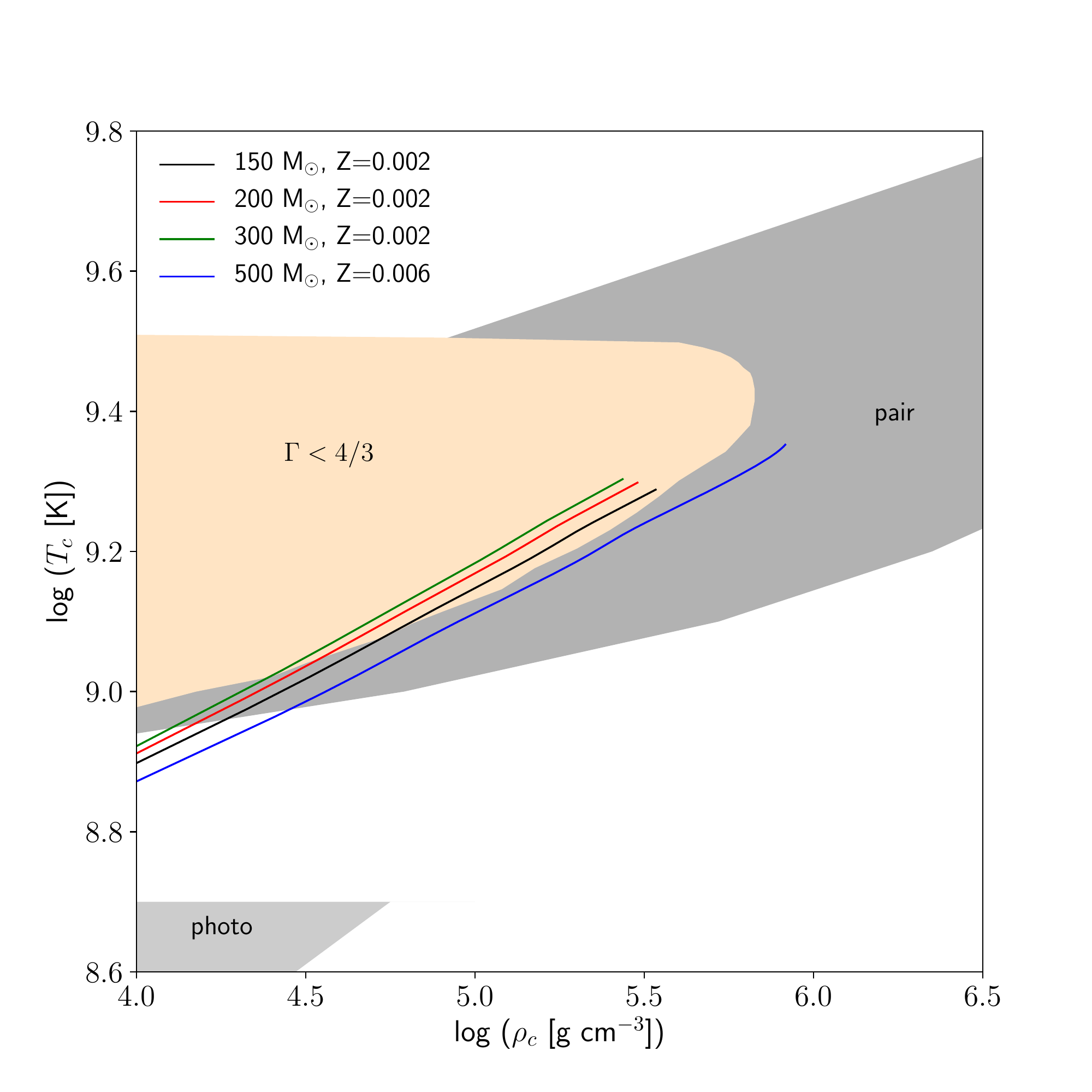}
    \caption{The evolution of central temperature and density of the models and the relationship with the neutrino process 
    regions. The  plot is the magnified version of Fig \ref{fig:neuprocess_general} where we constrained the central temperature  and  density regions to see 
    the tracks clearly. Shown in the plot is the instability region, the regions of pair production and photoneutrino processes.}
    \label{fig:neuprocess_magnified}
\end{figure}

We illustrate all the regions where the respective neutrino process is dominant over other processes shown in  Figure \ref{fig:neuprocess_general}. To highlight the impact of the thermal neutrino energy loss on the VMS models, we include the evolutionary tracks of the VMS models up to the O-burning at the end of the evolution. The evolutionary tracks and selected structure of the models are described in the next section. From the evolutionary tracks, we can see clearly the photoneutrino and plasmon decay processes  dominate the energy loss of very massive stars throughout the evolution.

\section{Results : Thermal Neutrino Energy Losses in Very Massive Stars}
\subsection{Input Physics of the Very Massive Stars}
\label{sec:physics} 

In our calculations, VMS models were calculated with the GENEC stellar evolution code by \cite{yusof2013evolution}. Full details of the 
very massive stars can be obtained from the same publication. It was suggested that for models that were predicted to end as a PISN are
three models from the SMC metalicity (Z=0.002) which are 150, 200 and 300 M$_{\odot}$ rotating models and  the LMC metallicity (Z=0.006), 
500 M$_{\odot}$  rotating model. We include the SMC 300 M$_\odot$ model to complete the grids and for this model, it is expected to 
end up as a core-collapse supernova (CCSN) or massive black hole (BH). The physics ingredients in calculating these models are the same as in \cite{ekstrom2012grids} and \cite{yusof2013evolution}. We summarise below some of the few important ones used in these models:

\begin{itemize}

    \item The initial abundances of mass fractions of $^1$H, $^3$He, $^4$He and metal (Z) are listed in 
    Table \ref{tab:initial_abundance}. The mixture of heavy elements, Z is taken from \cite{2005ASPC..336...25A} except for 
    the Ne abundance which was 
    adopted from \cite{2006ApJ...647L.143C} and the isotropic ratios are taken from \cite{2003M&PSA..38.5272L}.

    \item For the treatment of convection, the Schwarzschild criterion is used and a modest overshooting with an overshoot parameter, 
    $d_\text{over}/H_P$ = 0.10, is used for core hydrogen and helium burnings only.
 
    \item Mass loss affected strongly the evolution and fate of the PISN progenitors. Prescription by \cite{vink2001mass} is 
    used for the main sequence stars. For the domain not covered by \cite{vink2001mass}, we implement the prescription 
    by \cite{de1988mass} to the models with log $T_\mathrm{eff}> 3.7$ while for log $T_\mathrm{eff}\leq 3.7$, a linear fit is
    performed to the data from \cite{sylvester1998silicate} and \cite{1999A&A...351..559V}.
    \item The neutrino prescriptions are taken from \cite{itoh1989neutrino} and \cite{1996ApJS..102..411I}. 
    \item The treatment of rotation in the GENEC code has been described in \cite{2012RvMP...84...25M}.
    \item For the nuclear reaction rates, we implement the basic 21 nuclear isotopes in the nuclear network.

\end{itemize}

 Note that we do not include the $e^+e^-$  equation of state (EOS) in the GENEC code. This can be seen in Figure \ref{fig:neuprocess_magnified} where all the selected VMS models
 enter the instability region instead of stopping before entering the adiabatic region.  
 For the final fate estimation, we use 
 the carbon-oxygen core mass (M$_\text{CO}$) instead of the mass of the helium core. This estimate shows similar fate for stars with the 
 same mass \cite{heger2003massive} of the CO core found in various investigations of the early Universe \citep{1984ApJ...280..825B}.

\begin{table}
	\centering
	\caption{Initial abundances  of $^1$H, $^3$He, $^4$He and metal (Z) of the models. }
	\label{tab:initial_abundance}
	\begin{tabular}{lcccr} 
		\hline
		 & $^1$H & $^3$He &$^4$He & Z\\
		\hline
		LMC & 0.7381 &4.4247e-5 &0.2559 &0.006\\
		SMC & 0.7471 &4.247e-5 & 0.2508. &0.002\\
		\hline
	\end{tabular}
\end{table}

\subsection{Evolution of Central Temperature and Density}

\begin{figure}
	\includegraphics[width=\columnwidth]{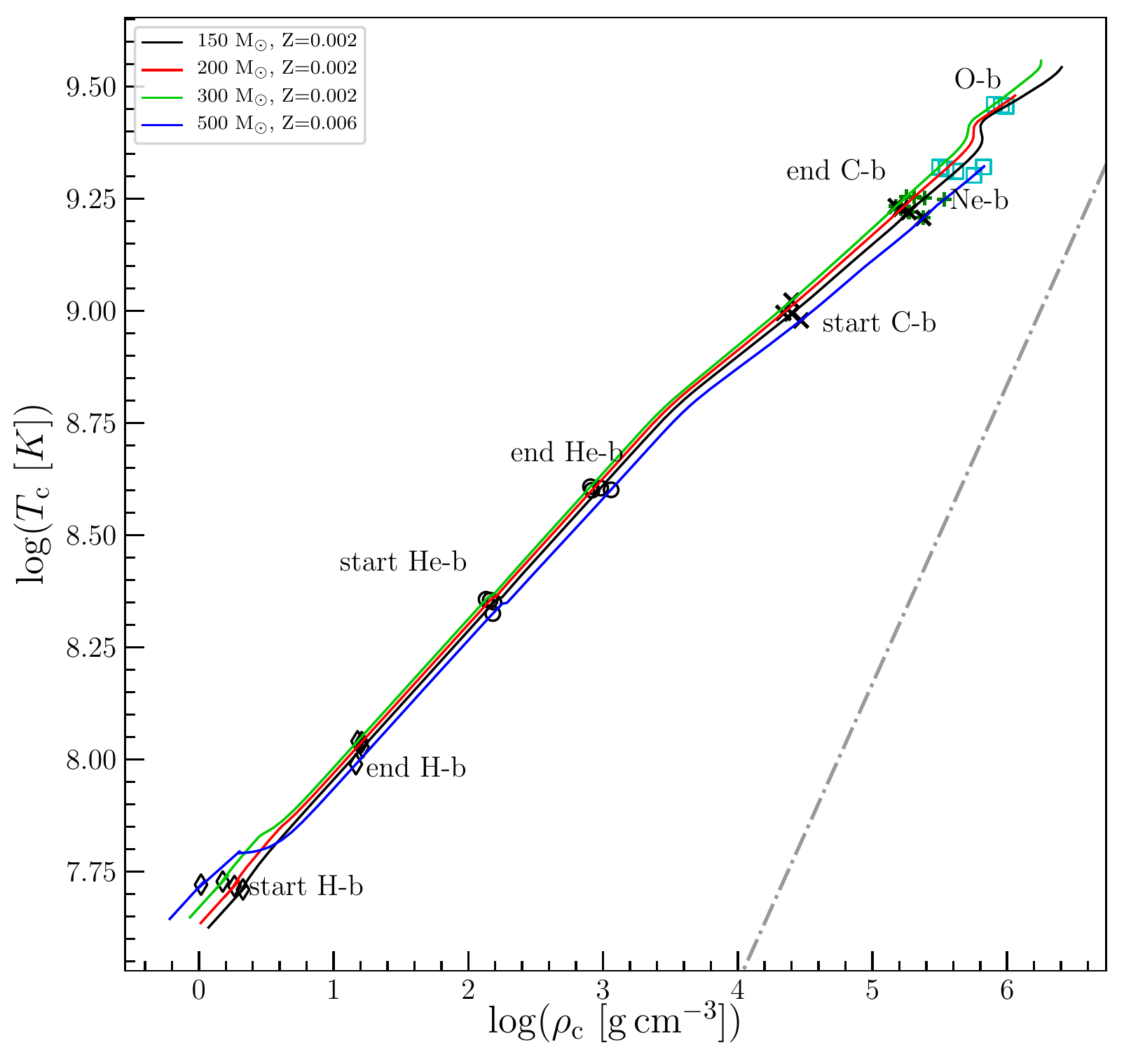}
    \caption{The evolution of central temperature and central density for 150, 200 and 300 M$_\odot$ 
    rotating SMC models and 500 M$_\odot$ rotating LMC model with the burning stages. The dotted line represents 
    the limit between non-degenerate and degenerate electron gas.}
    \label{fig:fulltcrhoc}
\end{figure}

Since the models chosen for this calculation focus on very massive stars, particularly the aforementioned progenitors that are expected to 
explode as PISN supernova, we carefully examine the temperature and density diagram to check the possibility of the 
stars entering the stability phase indicated by $\Gamma\geq 4/3$. 
Figure \ref{fig:fulltcrhoc} represents the central temperature-density ($T_c$-$\rho_c$) diagram with  the burning stages of all four models with the start and end of burning stages labelled in the graph. All four models selected in this work have had stopped  their evolution at the end of oxygen burning. Since we implement the mass loss mechanism in these models, 
our stars become a Wolf–Rayet (WR) star when the value of the surface hydrogen mass fraction, 
$X_s$ is below 0.3 and the effective temperature, $\log T_\mathrm{eff}$ would be greater 
than 0.4. Neutrino losses constitute a critical aspect in the stellar evolution of 
massive stars especially during the end of the He-burning phase 
\citep{woosley2002evolution}. Since neutrino losses are very sensitive with 
temperature and density, the rates of the neutrino losses are expected to increase when the temperature increases due to 
the heavier fuel that burn at the core of the 
stars. For very massive stars, this might be different since the initial 
temperature and density are higher than normal massive stars, thus we 
would expect to see neutrinos from thermal processes produced earlier than in massive stars 
since $T_c \approx 10^8$ K at the end of the H-burning phase (see Figure \ref{fig:fulltcrhoc}).

\begin{figure}
	\includegraphics[width=\columnwidth]{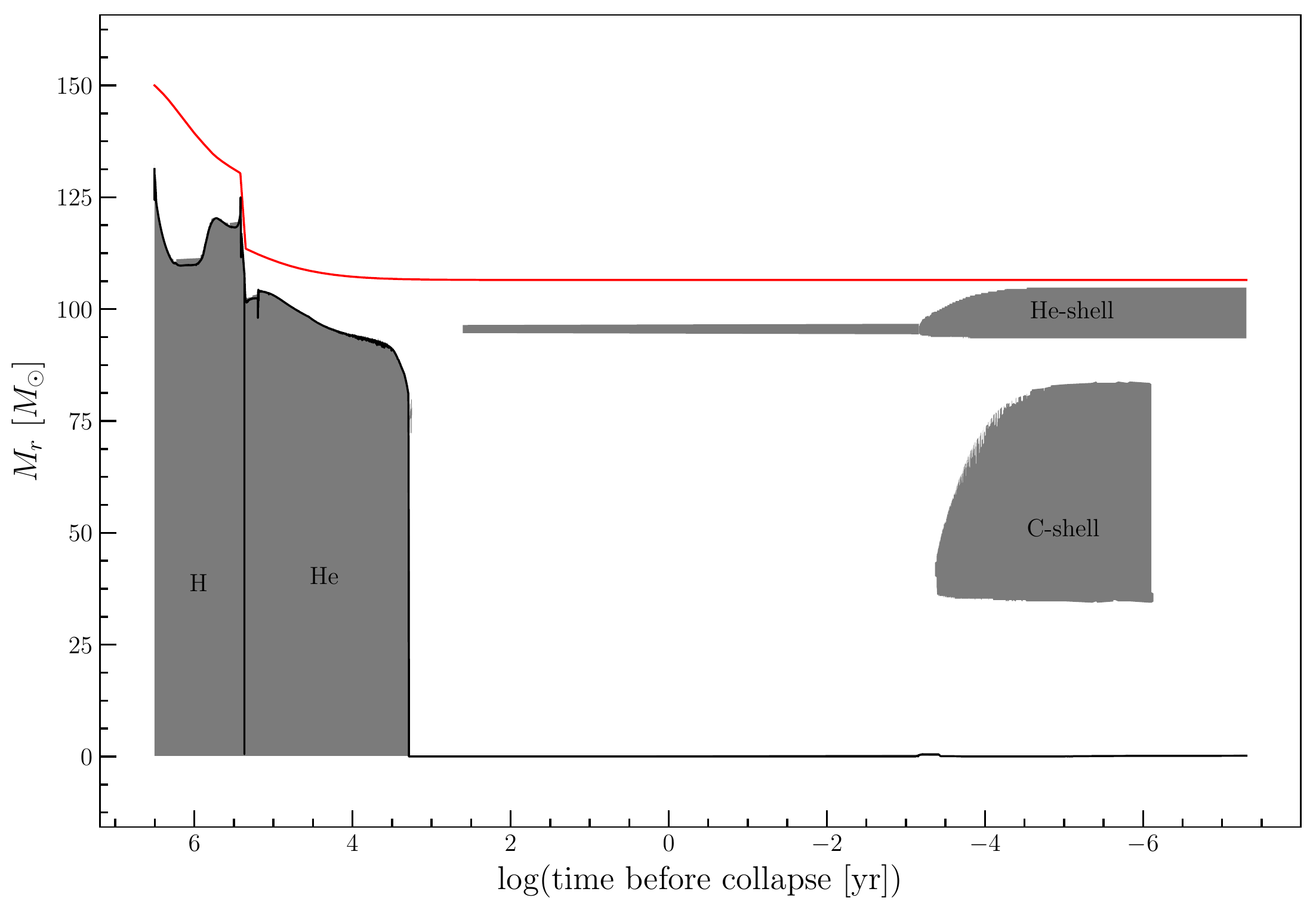}
    \caption{The Kippehahn diagram of 150 M$_\odot$ Z=0.002 SMC model w.r.t time before core collapse. The red line 
    represents the evolution of mass of the star and the ones in gray are the convective regions. The final mass of this model is around 106.5 M$_\odot$. Around 29\%  of mass loss is due to the stellar winds.}
    \label{fig:kip150}
\end{figure}

\begin{figure}
	\includegraphics[width=\columnwidth]{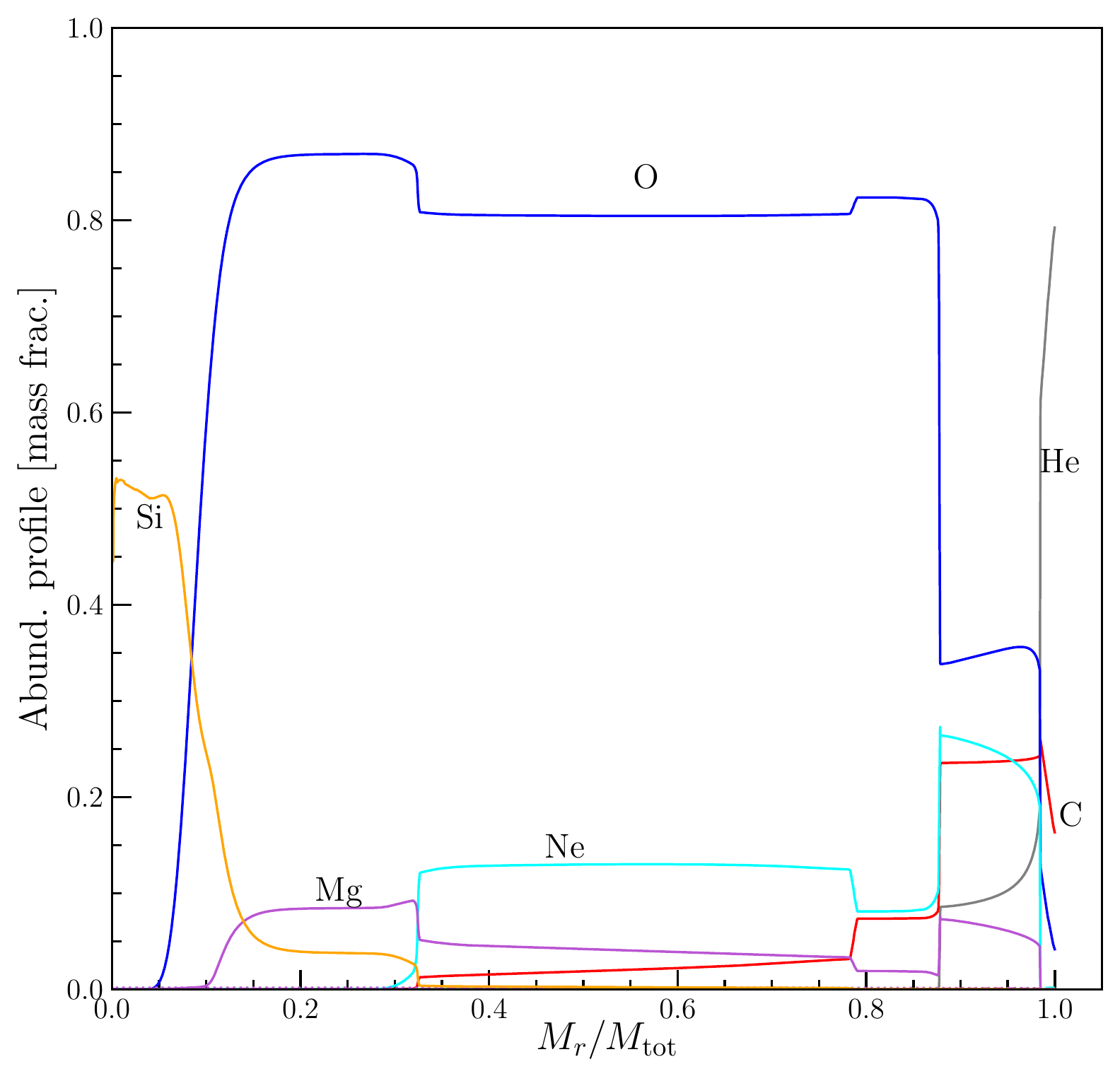}
    \caption{Abundance profile for 150 M$_\odot$ SMC at the final model. In this model, He and C are the most abundant at the 
    end of the evolution. }
    \label{fig:abun}
\end{figure}

Although the models were experiencing mass loss throughout their evolution and even when the rotating is included, the SMC models are able to retain their
mass during the main sequence.
This phenomenon can be observed in the Kippenhahn diagram as illustrated for 150 M$_\odot$ SMC model (see Figure \ref{fig:kip150}). We also presented the abundance profile for 150 M$_\odot$ with respect to the mass fraction in Figure \ref{fig:abun} at the oxygen core burning stage. In the abundance profile, our model contains 80\% of helium and around 20\% of carbon at the surface of the star. The hydrogen is totally depleted in our model. Since all the models are rotating models, the convective core size is higher than their corresponding non-rotating models convective core (for more detail discussion refer to \cite{yusof2013evolution}).

\begin{table}
	\centering
\caption{Initial mass,  the ratio of initial to critical velocity, metallicity, mass of carbon-oxygen core, final mass and fate of the models.	We defined CO core mass as mass 
coordinate at which sum of the carbon-oxygen mass fractions falls below 0.5. Data are taken from \protect\cite{yusof2013evolution} .}
	\label{tab:coremass}
	\begin{tabular}{lcccccr} 
		\hline
		M$_\text{ini}$ &$\frac{v_\mathrm{ini}}{v_\mathrm{crit}}$ &Z$_\odot$ &M$_\text{CO}$ &M$_\text{final}$ &Final product \\
		\hline
		150 & 0.4 &0.002 &93.468  &106.5  &PISN\\
		200 & 0.4 &0.002 &124.329 &129.2  &PISN\\
		300 & 0.4 &0.002 &134.869 &149.7  &BH\\
		500 & 0.4 &0.006 &73.145  &73.1   &PISN\\
		
		\hline
	\end{tabular}
\end{table}

\begin{figure*}
\centering
\begin{tabular}{cc}
\includegraphics[width=0.44\textwidth,clip=]{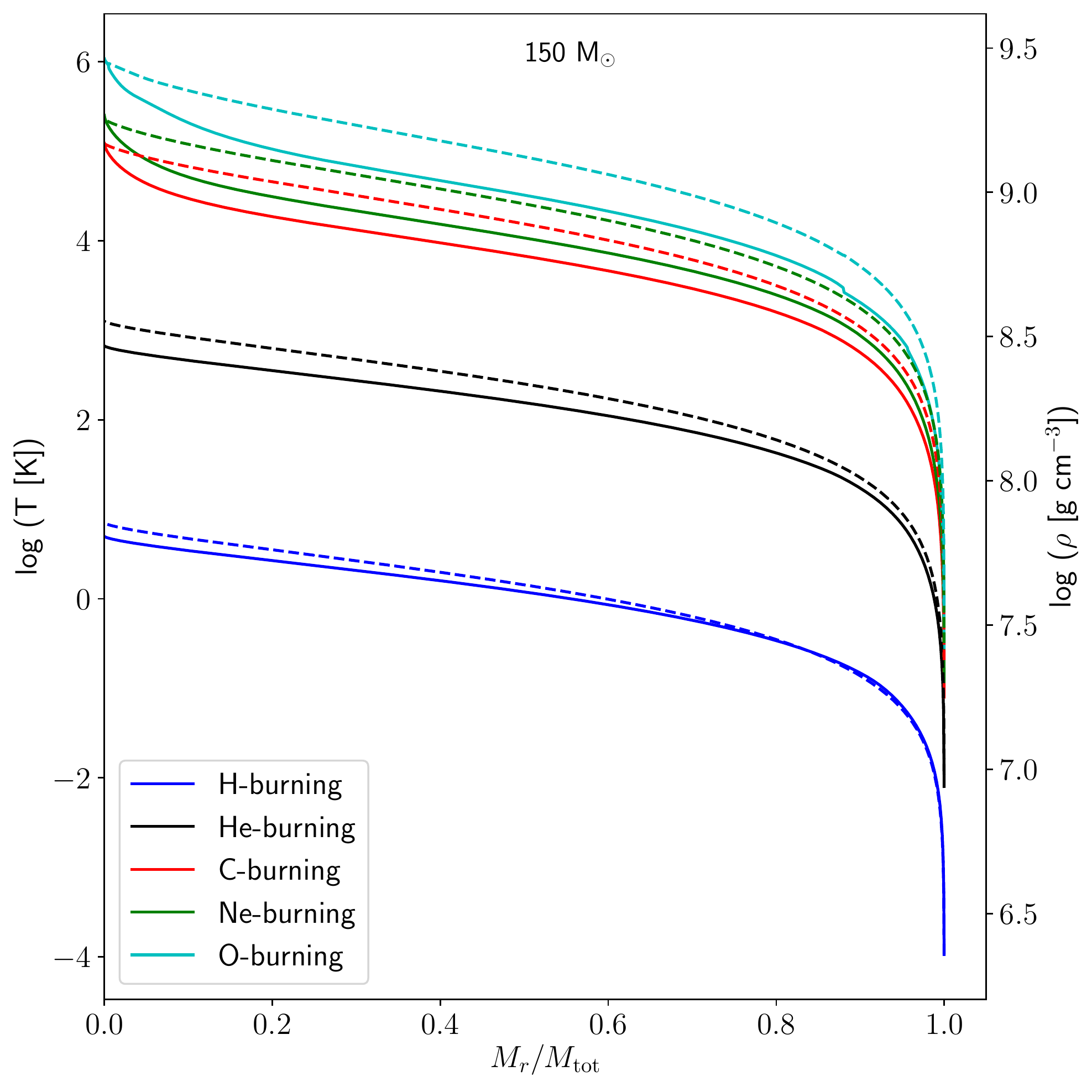} &
\includegraphics[width=0.44\textwidth,clip=]{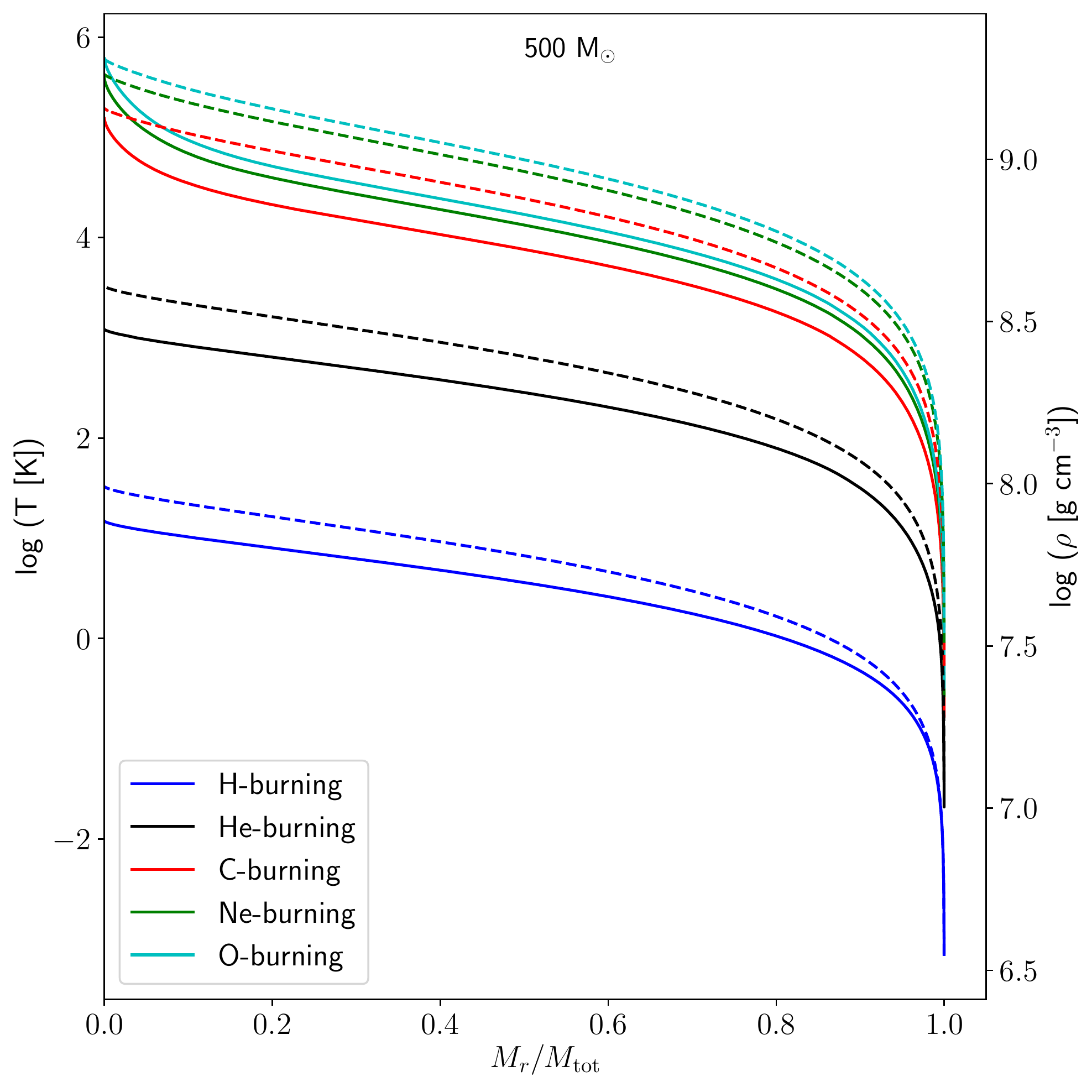} 

\end{tabular}
\caption{Temperature and density structure profiles with respect to the mass fraction at each burning stages for 150 M$_\odot$ and 500 M$_\odot$. The solid line represents the temperature vs mass fraction and the dotted line represents the density vs mass fraction.}\label{fig:burntrho}
\end{figure*}

\section{Result : Neutrino loss in Very Massive Stars}

For the neutrino energy loss calculation, we post-processed the stellar evolution data using neutrino energy loss fitting 
from \cite{1996ApJS..102..411I} and the references therein. Although the general prescription is already 
included in the GENEC code, but 
the output of the evolution of neutrino energy loss are not available in 
the grid. We have predicted the fate of very massive 
stars calculated in the models \citep{yusof2013evolution} and 150, 200 M$_\odot$ SMC and 500 M$_\odot$ LMC  are expected to 
become pair instability supernova while 300 M$_\odot$ end up as CCSN or BH based on the estimate from the 
carbon-oxygen core. We performed post-processed calculations based on \cite{1996ApJS..102..411I} to produce 
the details of the neutrino energy loss or emissivity from thermal processes, $Q_\nu$ for various neutrino processes. We also performed the 
luminosity calculation, $L_\nu$ for all models at the various nuclear burning stages. In this calculation, we do not include
the neutrino production from the $\beta^{\pm}$ decay and $e^\pm$ capture processes. 

\subsection{Energy Loss from Thermal Processes and Neutrino Luminosity}

For  150, 200, 300 M$_\odot$ SMC and 500 M$_\odot$ LMC models, we calculated and generated the total neutrino emissivity, $Q_\nu$ and neutrino luminosity, $L_\nu$ for
five thermal neutrino processes at the end of five different nuclear burning stages, hydrogen (H-burning), 
helium (He-burning), carbon (C-burning),
neon (Ne-burning) and oxygen (O-burning). The reason we choose the end of the burning stages is we want to investigate how much the neutrino energy loss is produced before the final carbon burning stage since for very massive stars, the temperature and density are 
high enough to produce neutrinos as early as during the H-burning stage. The temperature and density are taken from the structure
of the models, which are at the chosen burning stages where we took at that particular time step, the values of the temperature, density 
 and the isotropic nuclei from the center to the surface of the star. 

We plot the structure profile for temperature-density versus mass fraction for 150 M$_\odot$ Z=0.002 (SMC) and 500 M$_\odot$ Z=0.006 (LMC)  from center to the surface of the stars (see Figure \ref{fig:burntrho}) as an indicator for the cross-check for the prediction of neutrino energy loss. These two variables are important as input to test the neutrino energy loss. The major difference in the profiles can be seen at H-burning when 500 M$_\odot$ has higher central temperature and density compared to the less massive VMS model.

\begin{figure*}
\centering
\begin{tabular}{cc}
\includegraphics[width=0.4\textwidth,clip=]{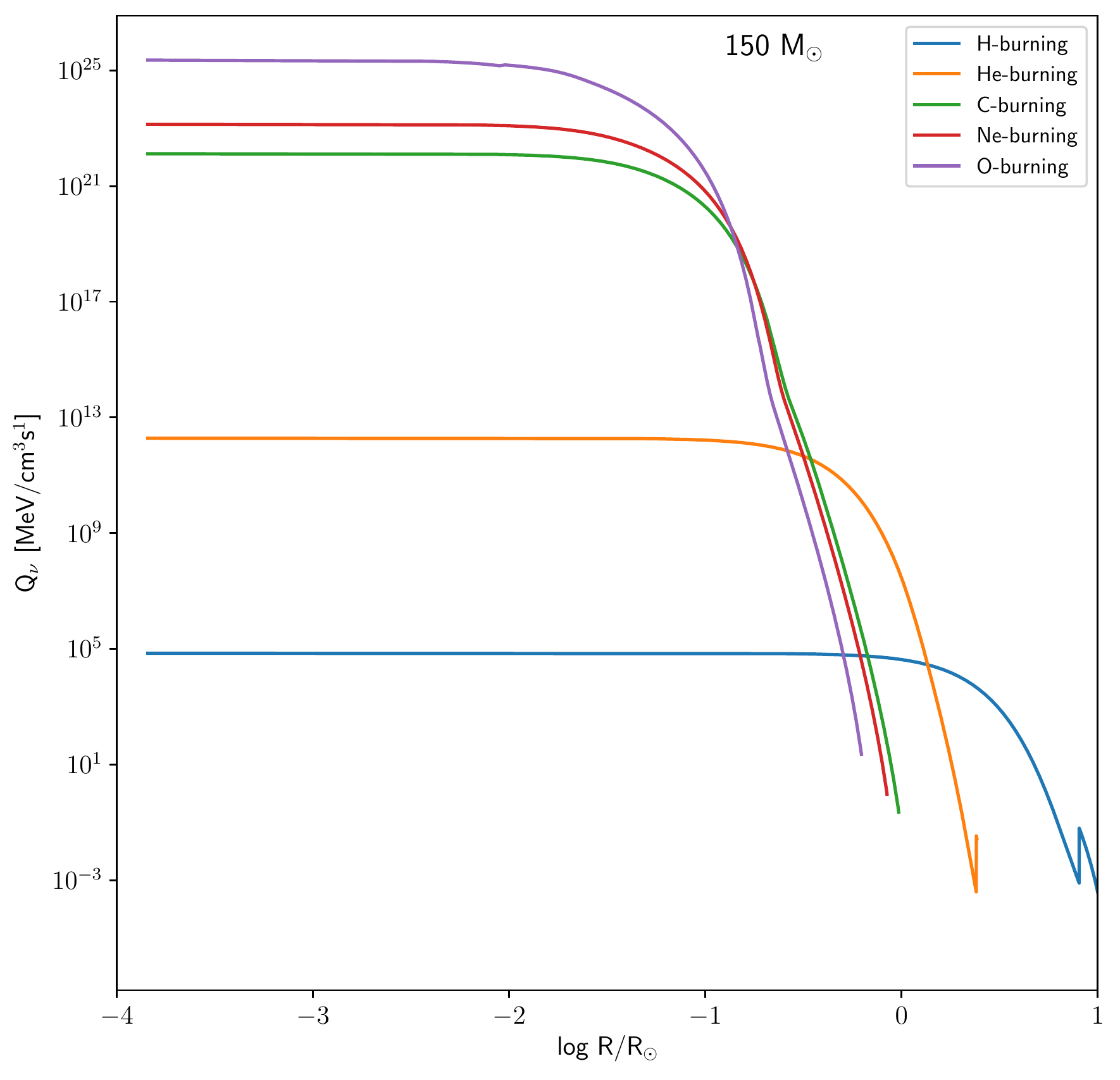} &
\includegraphics[width=0.4\textwidth,clip=]{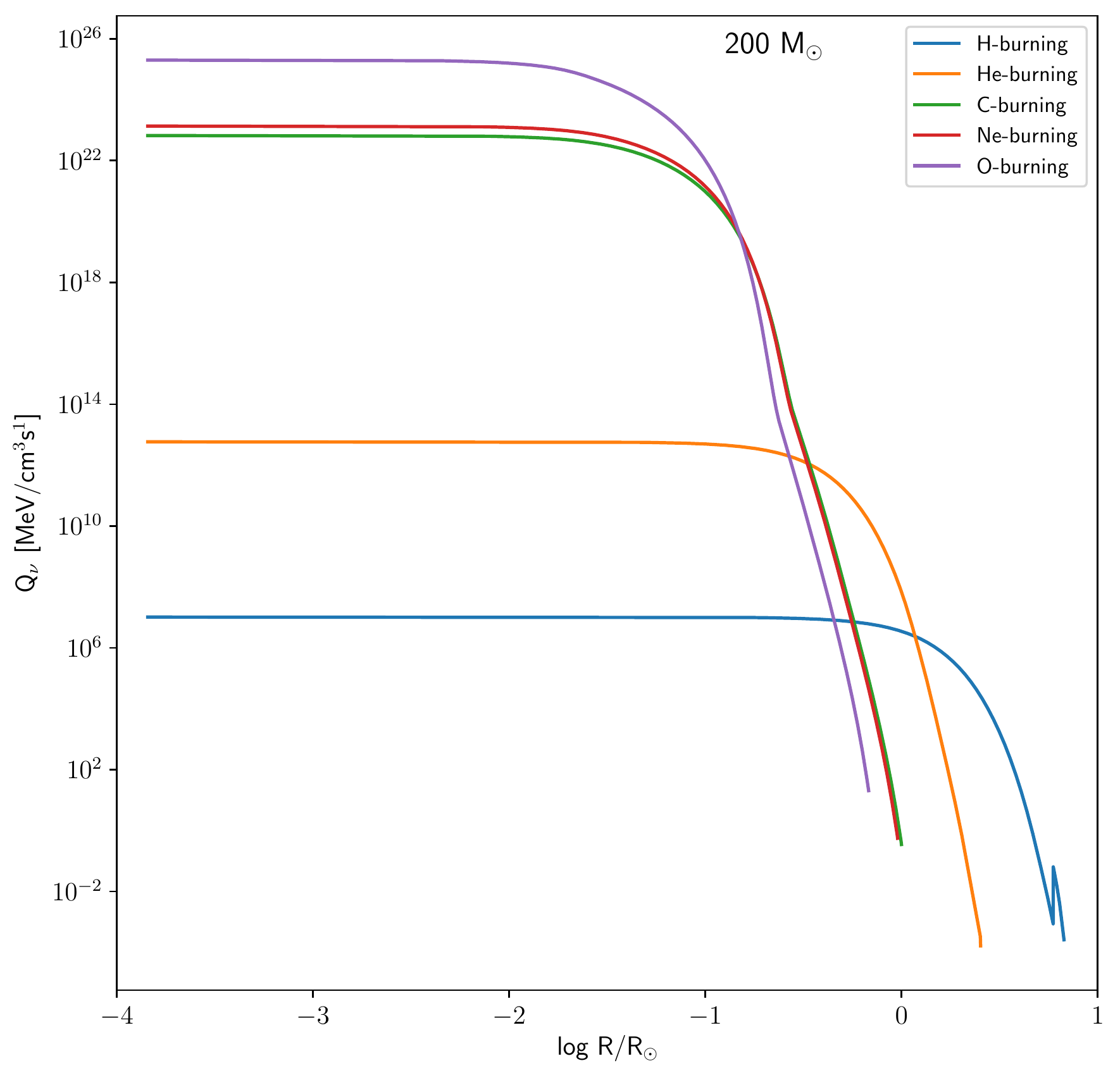} \\
\includegraphics[width=0.4\textwidth,clip=]{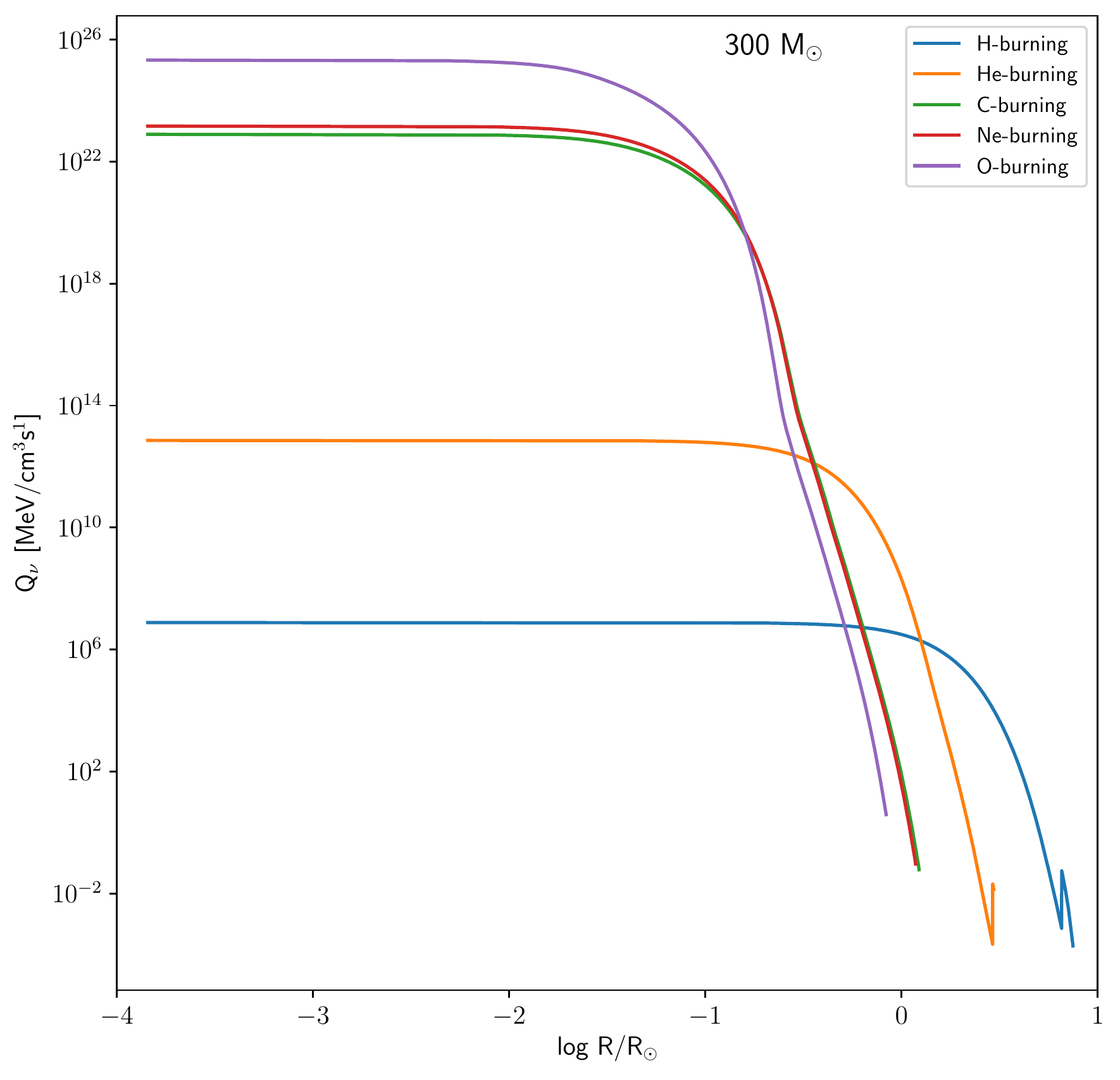} &
\includegraphics[width=0.4\textwidth,clip=]{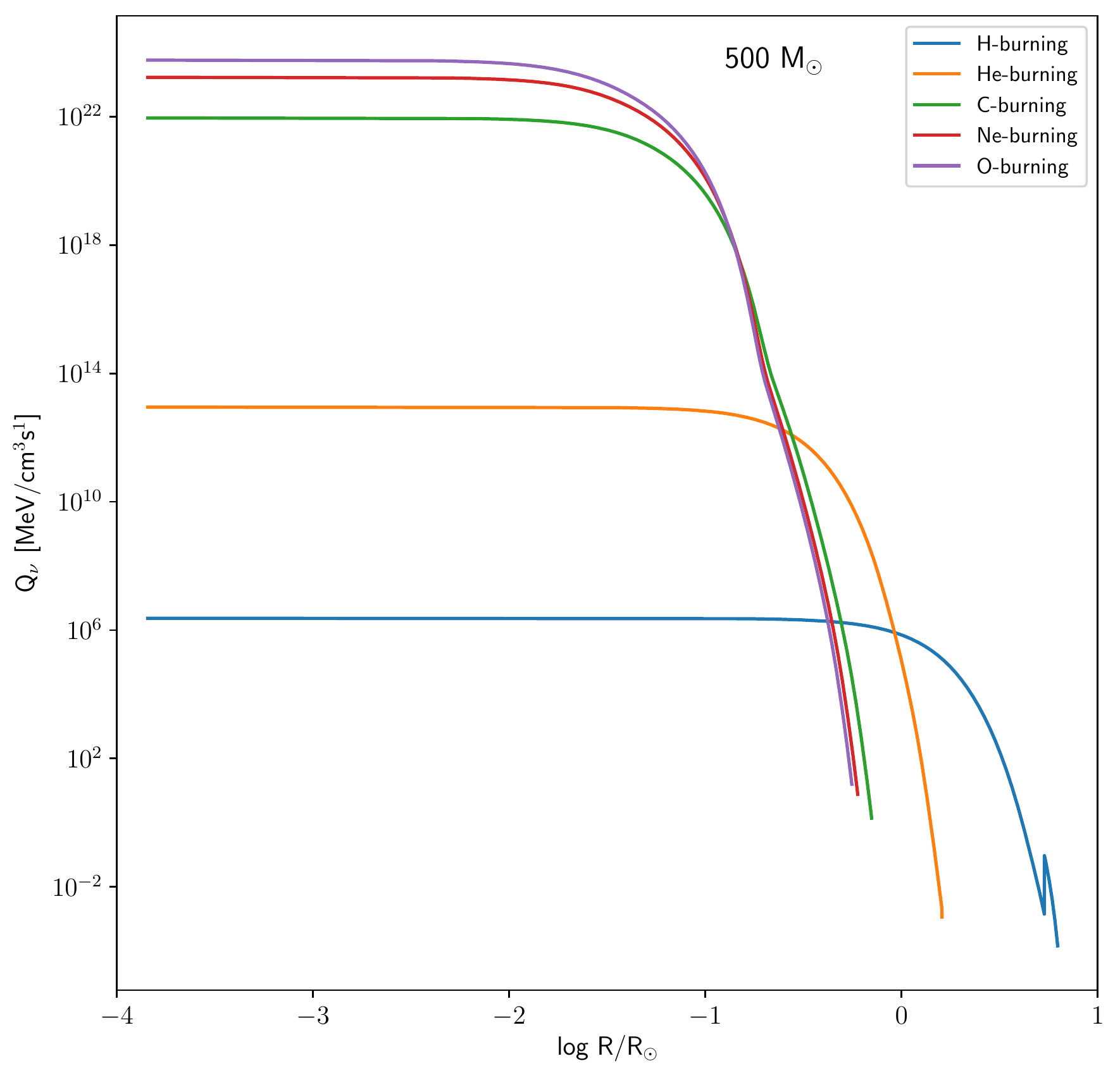} \\

\end{tabular}
\caption{Total neutrino emissivity for each burning stages for the progenitor models, $Q_\nu$ as a function of radial coordinate. }\label{fig:Qradius}
\end{figure*}

For very massive stars, their lifetimes are relatively shorter than its lighter siblings. More massive the star is, the temperature and density would be higher inducing faster nuclear reaction rates leading to shorter evolution time. For our selected progenitors, their lifetimes are around 2.3 - 3.0 Myrs. The H-burning (and total) lifetimes of VMS are lengthened by rotations as in lower mass stars \citep{yusof2013evolution}. As the stars go more massive, it evolves more rapidly where the lifetimes for each stage of nuclear burning depends on the progenitor mass. Neutrino energy loss, if not included in the evolution as early as helium burning stage, the carbon burning will start as soon as helium in the core is depleted \citep{rose1969neutrino}. Our reason to investigate the neutrino energy loss starting at the early until advanced burning stages is to see how sensitive are density and temperature dependent towards the production of thermal neutrinos since the initial temperature of VMS is relatively high compared to its less massive siblings.     
From our progenitor models, the lifetime of each burning stages for different masses are relatively homogeneous (see Table \ref{tab:schematic}). Small differences between metallcities are due to the different rate of mass loss at different metallicities.  Figure \ref{fig:Qradius} shows the radial distribution of $Q_\nu$ at each burning stages for each progenitor models. Most of the time, $Q_\nu$ is maximum at $\log R/R_\odot \approx -4$ which is located at the central region and start declining rapidly when the radius becomes larger (i.e towards the surface). The same pattern of the $Q_\nu$ is observed in massive stars progenitors \citep{patton2017presupernova} even the radius of their massive stars is smaller than our models. We note that there is a sharp discontinuity at the surface of the stars at hydrogen and helium burning phases which reflects the shell structure of the star. During H-burning, we observed the neutrino emissivity varies from one model to another due to the difference in the temperature range when the stars start to enter the main sequence. 

Since our models are pre-supernova models, at the end of O-burning, the total emissivity, $Q_\nu$ from our VMS models are three orders of magnitude higher than the massive star models except for 500 M$_\odot$ LMC model since it has the same magnitude of neutrino emissivity at the core of the star (see \cite{patton2017presupernova} Table 2). Stars with lower metalicity (in this case SMC) are more luminous with higher effective temperature. This is reflected in the central temperature where the location of 500 M$_\odot$ LMC track is at the most bottom in the $T_c-\rho_c$ diagram (refer to Figure \ref{fig:fulltcrhoc}).

\begin{figure}
\centering
\includegraphics[width=0.44\textwidth,clip=]{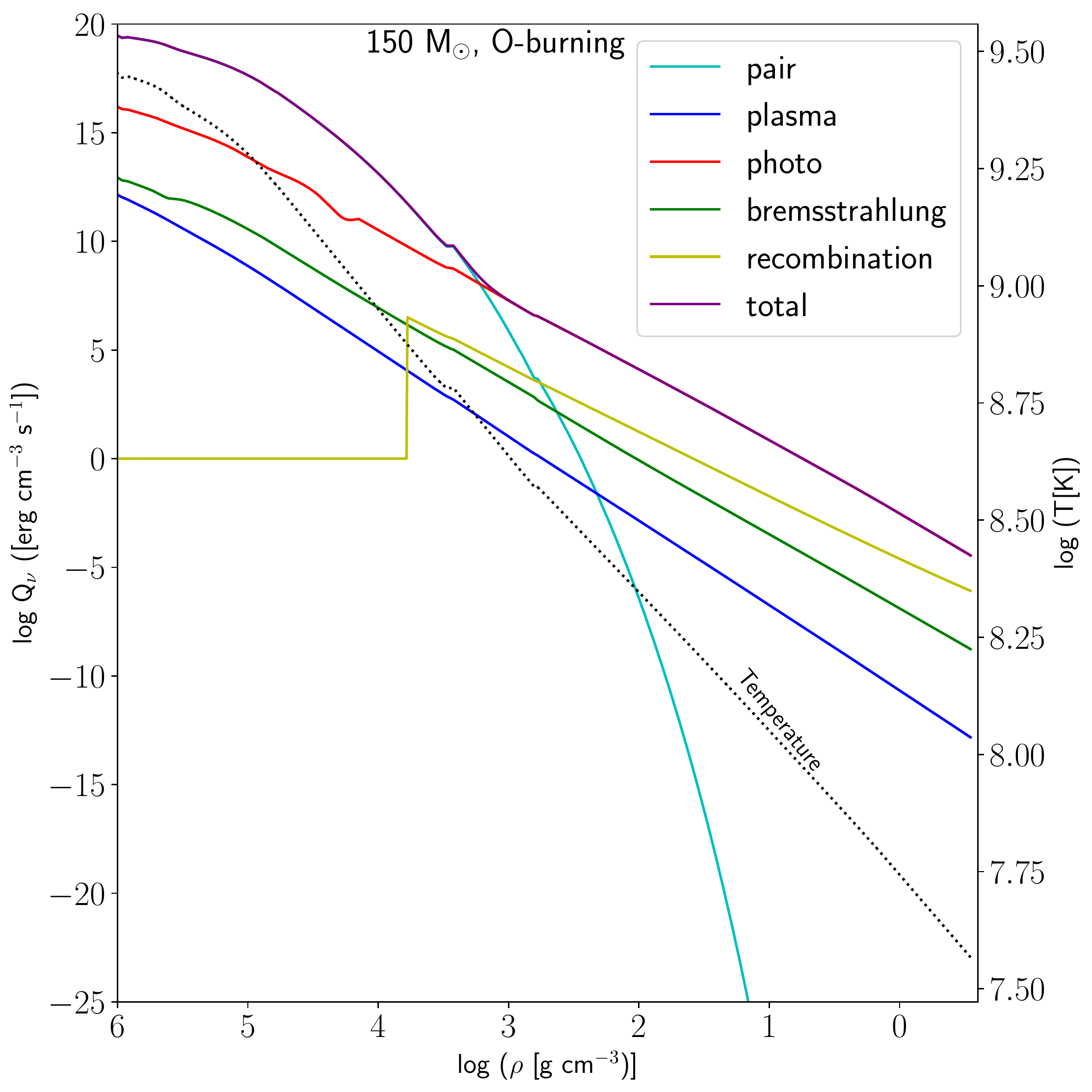} 

\caption{Comparison of energy loss, $Q_\nu$ as a function of temperature, $T$ and density $\rho$ 
for all thermal neutrino processes: plasma, bremsstrahlung, photo, pair and recombination at the structure profile of 150 M$_\odot$ O-burning stage. The total $Q_\nu$ is presented by purple line. The dotted line represents the relation between temperature and density.} 

\label{fig:QrhoT}
\end{figure}

In Figure \ref{fig:QrhoT}, we present the result for all neutrino processes (pair, plasma, photo, bremsstrahlung and recombination) and the comparison with the sum of all processes represented by the purple line for 150 M$_\odot$ using the structure profile at  O-burning stage. We particularly choose this stage since we expect our model to stop evolving due to the star is about to enter the PISN stage.  The grid points for the x-axis represent the density with range $10^{0.1}\mathrm{g cm^{-3}}<\rho <10^{7}\mathrm{g cm^{-3}}$ and temperature (right Figure \ref{fig:QrhoT}) of the model from surface to center which spans from $10^{7.5} \mathrm{K} < T< 10^{9.5} \mathrm{K}$. 
It is interesting to note that the pair neutrino process completely dominates the neutrino loss in the inner half of the star before it contributes the minimal towards the total loss after that and then photo-neutrino process takes over until the surface.  Active neutrino productions can be seen clearly at the core or centre of the star when the most dense and hot region located. In the hot dense core 
of very massive stars, when the temperature $T \gtrsim 10{^9}$ K, 
the electron-positron pairs are produced because they are in equilibrium with 
radiation. The interchanging of $e^-+e^++ \leftrightarrow{\gamma+\gamma}$ occur and this can be annihilated via weak interaction too, $e^-+e^+ \leftrightarrow{\nu+\bar{\nu}_{e}}$. This contributes to the pair annihilation process that is important in both massive and very massive stars. Degenerate and non-degenerate environments affect the thermal processes that occur in the VMS. For pair annihilation, this process declines as $\rho^{-1}$ in non-degenerate regions and causes the star to contract. In degenerate situations, the electron-positron pair creation suppresses the pair neutrino production and the energy loss rate of the neutrinos drops.

Photoneutrino from the analog of Compton scattering dominates at the non-degenerate electron gas at lower energy which can be describe as
 \begin{equation}
     \gamma + e^{-} \rightarrow \nu + \bar{\nu}+ e^{-}.
 \end{equation}
 Photoneutrino process plays an important role in removing the thermal energy from 
 degenerate core during the early stage of H and He nuclear burnings. During the 
 early stage of evolution the dominant process that involve is photoneutrino for 
 all models of VMS (see Figure \ref{fig:neuprocess_general}). This process is 
 severely suppressed at very high density because of the absence of unoccupied 
 electron states and pair production also similarly suppressed at very high density 
 and causes the pair neutrino emission greatly reduced \citep{festa1969neutrino}.

Based on the relation of neutrino emissivity and temperature-density region, Figure 
\ref{fig:neuprocess_general} shows that bremsstrahlung, plasma and recombination 
processes are the least important in neutrino production in very massive stars and 
hence the contribution to the neutrino energy loss. In our result, neutrinos 
produced from bremsstrahlung overwhelmed the neutrino production from plasma decay 
process. At the region where the electrons are strongly degenerate, plasma decay is 
suppressed and it leads to the dominance of bremsstrahlung neutrino emission 
\citep{guo2016spectra}.   

When a star becomes dense, photons in the dense gas have an effective mass and are termed as plasmons and can decay into neutrino-antineutrino pairs. This physical process creates the plasma neutrinos. In the plasma decay, photons and longitudinal plasma may decay into neutrino-antineutrino pairs $\gamma \rightarrow{\nu_{x}+\bar{\nu_{x}}}$ of all flavors $x$. As mentioned in Eq. ($3$) and Eq. ($4$), for the $\nu_{\mu}$ and $\nu_{\tau}$ spectra, the value of axial part contribution is very small. This suppresses these neutrino flavors.  

Recombination neutrino process starts  at $\rho \lesssim 10^{3.7}$ g cm$^{-3}$ and temperature $T\lesssim 10^{8.9}$ K and this happens at mass shell $\gtrsim 0.8$. This is consistent with the temperature-density region shown in Figure \ref{fig:fulltcrhoc} where this process is dominant. At values of density and temperature greater than these limits, electrons are relativistic and the recombination treatment becomes invalid \citep{1996ApJS..102..411I}. As this process also depends strongly on the heavy element abundances $\thicksim$ Z$^{14}$, oxygen being the most abundant element drops drastically at this mass shell (Figure \ref{fig:abun}) resulting in the reduction of the recombination neutrino emissivity towards the surface.

From Table \ref{tab:schematic}, we presented the summary for each burning stages in 
VMS including lifetime, central 
temperature, central density, neutrino luminosity for all thermal processes (plasma, bremsstrahlung, pair, photo and 
recombination) and the energy loss, $Q_\nu$. From our calculation, at the early 
burning stages (H and He), photoneutrino 
process contributed the most in the production of thermal neutrinos. The value of 
luminosity is  lower than the value 
predicted by \cite{odrzywolek2010neutrino} for H-burning phase since we only 
considered thermal processes and do not include 
any contribution from $\beta$ decay processes. However, the neutrino luminosity at 
He-burning phase from photoneutrino 
process and not from plasma process that contributed around $10^{39}$ erg/s which is higher from other previous studies (see 
for example \citep{odrzywolek2010neutrino}. This is due to VMS has higher density 
and temperature ($10^2$ gcm$^{-3}$ and $10^8$ K 
respectively) that falls within the photoneutrino region. This is also confirmed by our illustration in Figure 
\ref{fig:neuprocess_general}. Carbon burning phase is an important phase in the life of massive or very massive stars. During this phase, extreme temperature regime is required for C-burning and this induced the production of $e^+ e^-$ pairs from  
thermal distribution. These pairs can annihilate into $\nu\bar{\nu}$ pairs. At this stage, the central temperature of VMS 
is $T \sim 10^9$ K and this leads to the powerful neutrino energy loss in the stars. Neutrinos from thermal 
processes have overtaken nuclear energy in the energy production that make the star shift to a  $\nu$-cooled star. The 
burning process continues to the Ne and O-burning stages. At these two phases, the duration of the burning in very massive stars
are relatively very short,   $\sim 78$ and $\sim 26$ days respectively where the core of oxygen burning becomes a classical 
neutrino-cooled stage. We also observed like in the C-burning phase, the pair neutrino is the most dominant process at Ne and O-burning phases given the high temperature and density at these phases. For our progenitors models, all models fall within 
the PISN explosion based on the estimate of M$_\mathrm{CO}$ (see \cite{yusof2013evolution} for details). From our luminosity 
calculation, the orders of magnitude for all neutrino processes are consistent with calculation done by \cite{wright2017neutrino} and \cite{leung2020pulsational} although their progenitor models have lower metallicity than our models. 
During the neon/oxygen burning phase, the VMS will experience instability due to the production of $e^+ e^-$ pairs from 
photons. The creation of $e^+ e^-$ pairs in the oxygen rich core soften the equation of states, leading to further 
contraction and this runaway collapse is predicted to produce a powerful explosion, with energy $10^{53}$ erg, that would 
break entirely the structure of stars without remnants \citep{kasen2011pair,whalen2014pair,gilmer2017pair}.

\subsection{Neutrino Luminosity}

The stars are assumed to be in local thermodynamic equilibrium and thus the time derivative would vanish 
leaving the neutrino luminosity, $L_\nu$ production described by the differential equation
\begin{equation}
\frac{dL_\nu}{dM_r}=\epsilon_\nu(\rho,T)
\end{equation}
where $M_r$ is the mass shell at the radius $r$ of the star. The luminosity differential equation is integrated over 
the models to obtain the total luminosity at the surface of the stars. We use the neutrino emissivity $Q_\nu$ to 
calculate the rate of the neutrino energy $\epsilon_\nu(\rho,T)$ released per unit stellar matter from the thermal 
processes. This integration is done for all burning stages of each VMS.

\begin{figure}
	\includegraphics[width=\columnwidth]{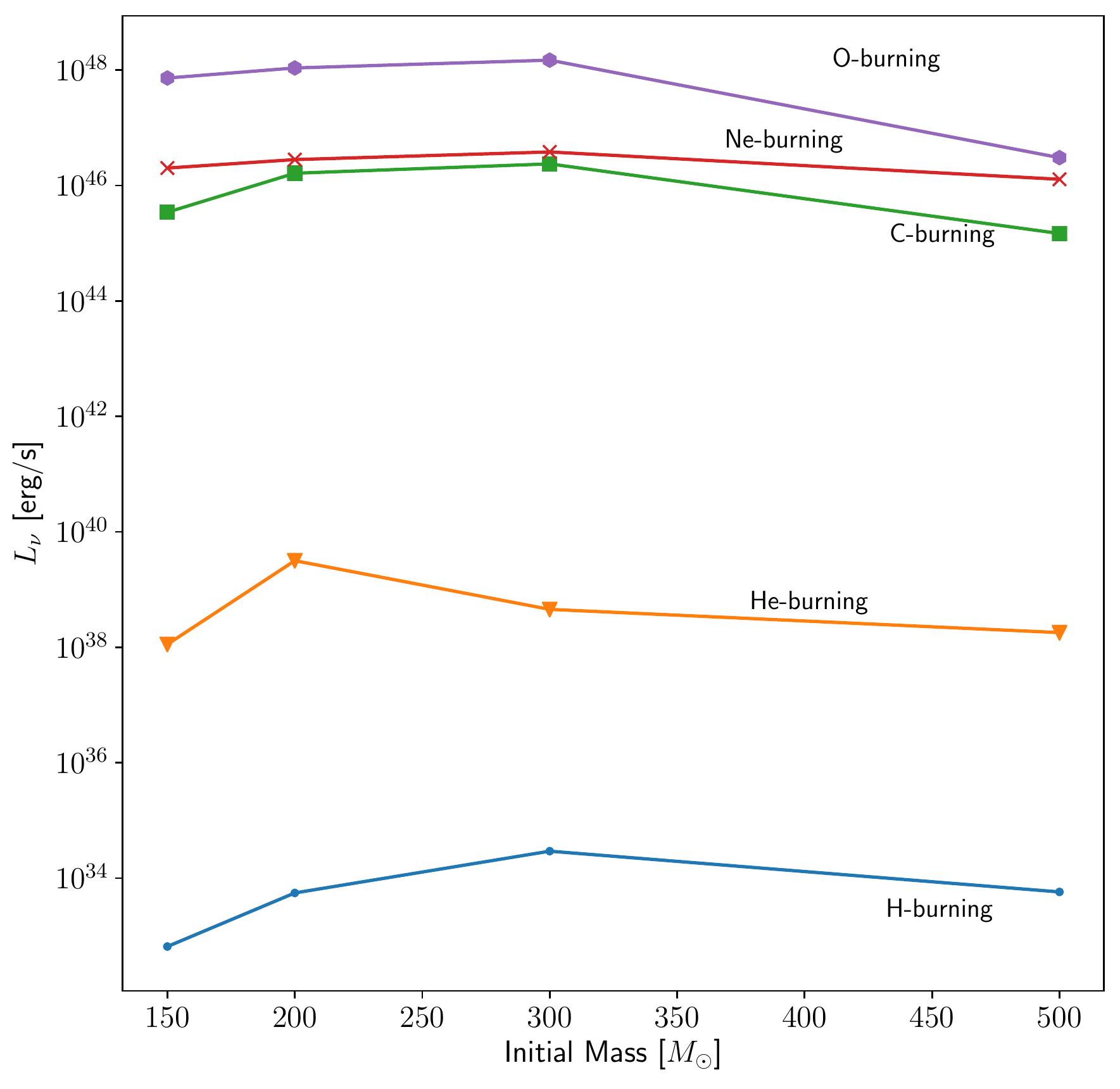}
    \caption{Relation between neutrino luminosity, $L_\nu$ with initial mass of the VMS models. For different initial masses, the luminosities are consistent throughout the burning stages. }
    \label{fig:luminositymodel}
\end{figure}

We presented the total luminosity $L_\nu$ of neutrino production at different burning stages with respect with the 
initial mass in Figure \ref{fig:luminositymodel}. From the graph, we verify that neutrino productions from thermal 
processes increased rapidly when the stars start to experience the C-burning phase. At the O-burning stage, where the 
luminosity $\approx 10^{51}$ erg/s, we found that the models at SMC metallicty have higher luminosity compared to the 
model at the LMC metallicity even when the mass is higher in LMC. This result is consistent with the total luminosity 
calculated for pulsation pair instability supernovae (PPISN) by \cite{leung2020pulsational} and PISN 
\citep{wright2017neutrino}. For 500 M$_\odot$ model, their lifetime span is shorter than the lighter VMS SMC models 
and less hotter at O-burning phase due to initial temperature difference during ZAMS between these two metallicities. 
Although there are work (see \cite{moriya2014mass}) that show mass loss of massive stars near Eddington luminosity 
has impacted on the neutrino luminosity where mass loss has in fact increased the neutrino luminosity; this impact 
didn't effect much for VMS stars.

\begin{figure}
	\includegraphics[width=\columnwidth]{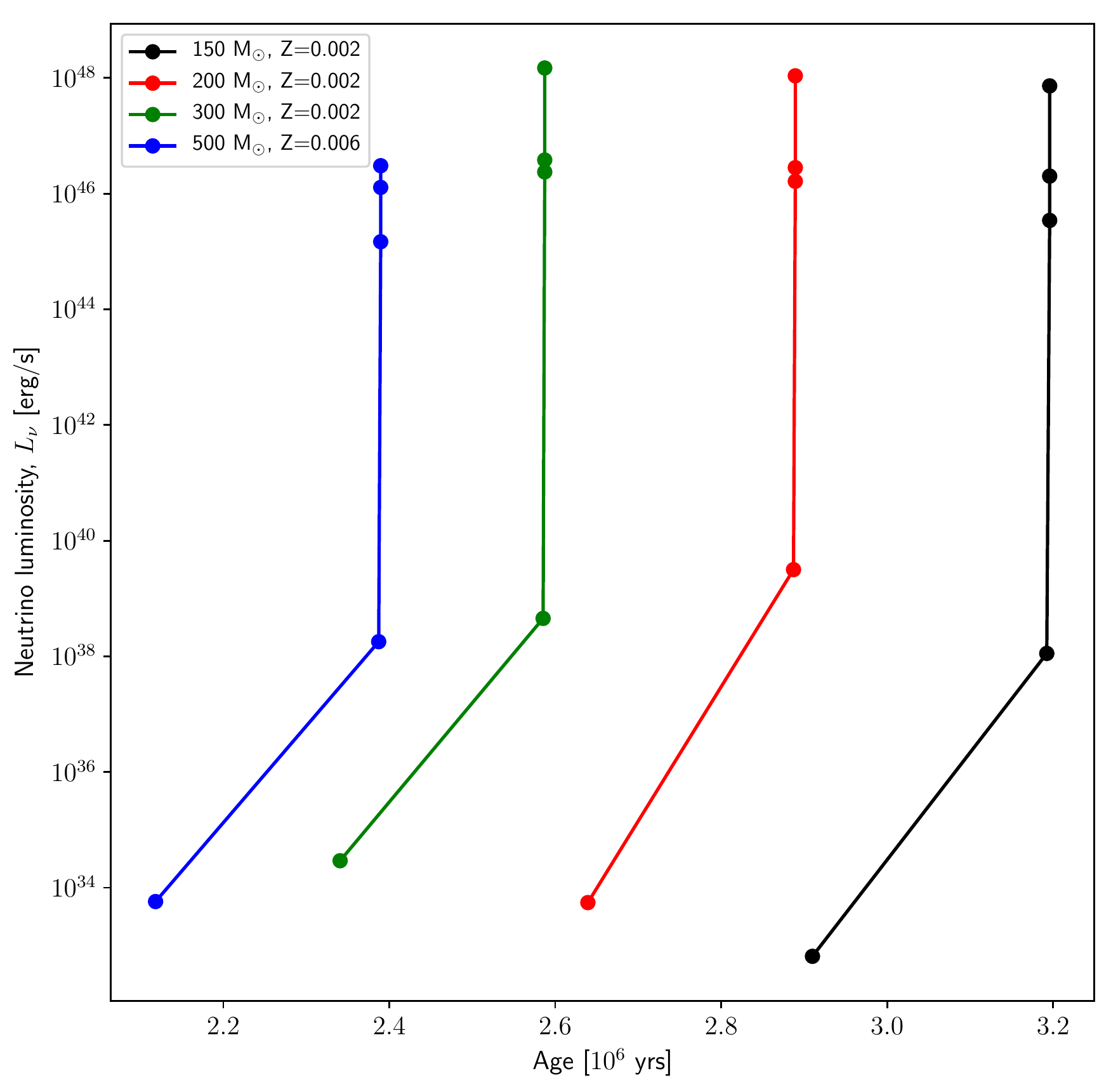}
    \caption{Neutrino luminosity, $L_\nu$ of progenitor models with respect to the age of the stars. Each point corresponds to the lifetime of H, He, C, Ne and O-burning phases.}
    \label{fig:evoluminosity}
\end{figure}

The evolution of neutrino luminosity $L_\nu$ with age for each nuclear burning stages is shown in Figure 
\ref{fig:evoluminosity}. From the graph, the evolution of neutrino luminosity is consistent with mass at the same 
metallicity. Comparing with the neutrino luminosity of 20 M$_\odot$ ($L_\nu=7.4 \times 10^{39}$ erg/s) calculated by 
\cite{odrzywolek2004detection}, $L_\nu$ for the 150 M$_\odot$ model ($L_\nu\sim 10^{45}$ erg/s) is six orders of 
magnitude larger at the C-burning stage meanwhile at O-burning, the 150 M$_\odot$ neutrino luminosity ($L_\nu\sim 
10^{45}$ erg/s) is three orders of magnitude larger than the 20 M$_\odot$ ($L_\nu\sim 10^{43}$ erg/s) at the same 
stage. Our neutrino luminosity calculation is in agreement with the total PISN neutrino luminosity for 250 M$_\odot$, 
$Z=10^{-3}$ \citep{wright2017neutrino} where the progenitor model was generated from the same stellar evolution code. 
It is well known that photon luminosity is always greater than neutrino luminosity of stars at the main sequence 
\citep{masevich1965neutrino,shi2020neutrino,farag2020stellar}. It would be an interesting test to estimate the 
difference between photon luminosity, $L$ with neutrino luminosity for each stages. 
In Table \ref{tab:lumphoton}, we demonstrate the comparison between these two different luminosities. For photon 
luminosity, photon energy is released by $\sim 10^{40}$ erg/s  throughout its evolution. For neutrino luminosity, the 
value is smaller at the main sequence which confirms our hypothesis with a difference around seven orders of magnitude lower than the photon luminosity. Even in He-burning, photon luminosity is higher since energy generated from the 
stars is dominated by nuclear reactions. The neutrino luminosity increases rapidly when its enter the neutrino cooling phase which is C-burning stage and it has overtaken the photon luminosity by six orders of magnitude indicating that 
large number of neutrinos are escaping from the stars.

 From our result, pair annihilation process is dominant during the 
 advanced stages of of evolution, C-burning, Ne-burning and O-burning for 
 $150$ M$_\odot$, $200$ M$_\odot$ and $300$ M$_\odot$ models as shown in Figure \ref{fig:neuprocess_general}. Although we investigated the emissivity of all types of neutrinos from the progenitor phase, it is beyond the scope of this paper to discriminate the contribution of each neutrino flavor on the emissivity.

\section{Conclusions}
We have post-processed the data from selected VMS grids; 150, 200 and 300 M$_\odot$ SMC and 500 M$_\odot$ LMC models particularly in the interest of PISN progenitors produced by the GENEC code to demonstrate the neutrino production from all the possible thermal neutrino processes. This initial ground work is important since we have very little knowledge and understanding about neutrino astronomy in this particular mass range (>100 M$_\odot$). We have extracted the neutrino energy loss and luminosity using selected models of pre-supernova progenitors. We found that at early life of the stars, i.e in H and He-burning stages, the most dominant thermal neutrino process is photoneutrino process. This prediction is different for massive stars where it was assumed to be dominated by plasma process at He-burning phase. This is because in massive stars, the core temperature is around 10$^8$ K while for VMS the temperature is around 10$^9$ K as thermal neutrino processes are very sensitive to the changes in temperature and density at the center of the stars. For the neutrino-cooled stars (starting from C-burning phase), the pair neutrino takes over in the neutrino production in the star. This is similar to the neutrino energy loss in massive star models \citep{woosley2002evolution, odrzywolek2004detection, patton2017neutrinos}.

Although \cite{wright2017neutrino} has performed neutrino signature calculation from simulation of PISN, the approach of the calculation is different from this work where we have considered the production throughout each nuclear burning stages in the very massive stars models but in the calculation of $Q_\nu$, both our calculation and \cite{wright2017neutrino} are still within agreeable range despite the PISN progenitors are different in mass and metallicity. Nevertheless, our estimate for neutrino emission is agreeable with \cite{wright2017neutrino} however their models are lower in metallicity (Z=10$^{-3}$).
At the O-burning stage, the neutrino luminosity $\approx 10^{47-48}$ erg/s depending on their initial mass and metallicity which is slightly higher than luminosity from massive stars. We conclude that the temperature and density of stars play an important role in determining the neutrino emission where neutrino loss is very sensitive with these two parameters. With metallicity, the initial temperature at ZAMS is shifted according to either the stars are metal-rich or metal poor stars. Our intial work could give some background work in providing more information about the possibility of detecting neutrinos from very luminous supernovae like PISN.
We also have analyzed one progenitor of black hole, i.e. 300 M$_\odot$ SMC model and it gives a significant neutrino emission and luminosity. This could provide information on a possible mechanism to detect a massive black hole by measuring the neutrino flux after the very massive stars collapse directly to become one.

The same models have been used to determine the existence of very massive stars within SMC and LMC, which are the R136a1 and NGC3603 \citep{crowther2010r136} which are located within a distance of $d=49.97$ kpc. This work may shed light on the possibility of using the neutrinos to detect candidates for pair instability supernova in our local universe. The next step to further this work, we plan to calculate in more detail the neutrino spectrum, flux and to include the impact of neutrino oscillations in the neutrino calculations at the pre-supernova stage that could give more information about the prediction of the future neutrino detection.

\begin{landscape}
\begin{table}
	\centering
	\caption{Schematic tabulation of neutrino emission in 150, 200,
	300 M$_\odot$ Z=0.002 and 500 M$_\odot$ Z=0.006
	according to the burning stages, 
	lifetime (in  years), surface 
	temperature and surface density, luminosity of plasma, bremsstrahlung, pair, photo and recombination process
	total neutrino luminosity, neutrino 
	energy loss at the surface  of these VMS at each 
	burning stages.}
	\label{tab:schematic}
	
	\begin{tabular}{ccccccccccccccc} 
		\hline
	   Mass &Z &Stage &Lifetime  &log T$_c$  &log $\rho_c$  &$L_{\nu_\mathrm{plasma}}$  &$L_{\nu_\mathrm{brems}}$  &$L_{\nu_\mathrm{pair}}$ &$L_{\nu_\mathrm{photo}}$ &$L_{\nu_\mathrm{recomb}}$   &$L_\nu$ &$Q_\nu$   \\
	                & &      &          &  (K)       & (g cm$^{-3}$)  &(erg s$^{-1}$) &(erg s$^{-1}$) &(erg s$^{-1}$) &(erg s$^{-1}$) &(erg s$^{-1}$)         &(erg s$^{-1}$)     &(MeV/cm$^3$s) \\
		\hline
		150  &0.002 &H-burn  &2.194E+06 &7.854512  &0.698107 &0.299230E+26       &0.435648E+29       &0.141107E-24   &0.626608E+33   &0.258618E+32       &0.652514E+33   &0.702676E+05 \\
		     &      &He-burn &2.778E+05 &8.553730  &2.826858    &0.523869E+32       &0.253045E+35       &0.581330E+34   &0.111905E+39   &0.437396E+36       &0.112373E+39  &0.191335E+13   \\
		     &      &C-burn  &3.474E+03 &9.170006  &5.087911 &0.216955E+38       &0.887269E+39       &0.344242E+46   &0.220568E+43   &0.247714E+38       &0.344462E+46   &0.131956E+23  \\
		     &      &Ne-burn &8.958E-03 &9.254826  &5.401279 &0.127393E+39       &0.385343E+40       &0.200804E+47   &0.800899E+43   &0.198892E+38       &0.200884E+47   &0.137796E+24       \\
	         &      &O-burn  &2.937E-03 &9.463448  &6.034956  &0.677010E+40       &0.920095E+41      & 0.728408E+48   &0.203862E+45   &0.101773E+38       &0.728612E+48   &0.227414E+26   \\
		\hline
		200  &0.002 &H-burn  &2.639E+06  &8.061185 &1.277386  &0.201985E+28       &0.161054E+31       &0.691116E+03   &0.289198E+35   &0.458385E+33       &0.293798E+35     &0.102148E+08\\
		     &      &He-burn &2.478E+05  &8.604531 &2.934564  &0.144723E+33       &0.658381E+35       &0.370875E+36   &0.320197E+39   &0.956115E+36       &0.321590E+39     &0.582921E+13\\
		     &      &C-burn  &2.158E+03  &9.227954 &5.213010  &0.645715E+38       &0.262817E+40       &0.163213E+47   &0.669776E+43   &0.305013E+38       &0.163280E+47    &0.668109E+23\\
		     &      &Ne-burn &1.389E-03  &9.254484 &5.311696  &0.112158E+39       &0.418293E+40       &0.281254E+47   &0.973757E+43   &0.285844E+38       &0.281351E+47    &0.136524E+24  \\
	         &      &O-burn  &2.473E-03  &9.457851 &5.957464 &0.591853E+40       &0.108859E+42       &0.108805E+49   &0.253695E+45   &0.143258E+38       &0.108831E+49   &0.200181E+26  \\
	     \hline
		300 &0.002 &H-burn  &2.341E+06 &8.053011  &1.216090 &0.165469E+28       &0.148039E+31       &0.118875E+03   &0.288807E+35   &0.426979E+33       &0.293092E+35   &0.766218E+07 \\
		    &      &He-burn &2.446E+05 &8.616531  &2.935627 &0.180798E+33       &0.861075E+35       &0.107686E+37   &0.450886E+39   &0.118894E+37       &0.453238E+39   &0.720756E+13 \\
		    &      &C-burn  &1.975E+03 &9.234097  &5.177245 &0.684084E+38 &0.314259E+40       &0.237003E+47   &0.844625E+43   &0.391974E+38       &0.237088E+47  &0.789035E+23 \\
		    &      &Ne-burn &1.004E-03 &9.257358  &5.263301 &0.110770E+39       &0.473213E+40       &0.381659E+47   &0.117437E+44   &0.363485E+38       &0.381777E+47     &0.147287E+24   \\
	        &      &O-burn  &2.085E-03 &9.460846  &5.908152 &0.592012E+40       &0.128680E+42       &0.148673E+49   &0.308911E+45   &0.181276E+38       &0.148704E+49  &0.214259E+26   \\
	     \hline
		500 &0.006 &H-burn  &2.112E+06 &7.991346 &1.172134 &0.521307E+27       &0.418823E+30       &0.448855E-05   &0.561276E+34   &0.153558E+33       &0.576674E+34    &0.235009E+07\\
		    &      &He-burn &2.688E+05 &8.609029 &3.084778 &0.792509E+32       &0.367477E+35       &0.939259E+35   &0.178784E+39   &0.543441E+36       &0.179458E+39   &0.889445E+13 \\
		    &      &C-burn  &2.428E+03 &9.157241 &5.187825 &0.208108E+38       &0.646740E+39       &0.147319E+46   &0.126753E+43   &0.136312E+38       &0.147446E+46  &0.911658E+22 \\
		    &      &Ne-burn &1.585E-02 &9.262511  &5.593359 &0.194248E+39       &0.392525E+40       &0.128107E+47   &0.633926E+43   &0.100716E+38       &0.128171E+47     &0.168462E+24   \\
	        &      &O-burn  &1.679E-03 &9.310245 &5.785761 &0.532323E+39       &0.878894E+40       &0.303784E+47   &0.129561E+44   &0.865246E+37      & 0.303914E+47  &0.578410E+24  \\
	     \hline
	\end{tabular}

\end{table}
	\end{landscape}

\begin{table}
	\centering
	\caption{The neutrino luminosity and photon luminosity in 150, 200,
	300 M$_\odot$ Z=0.002 and 500 M$_\odot$ Z=0.006
	according to the burning stages, 
	lifetime (in  years)
	 at the surface  of these VMS at each 
	burning stages.}
	\label{tab:lumphoton}
	
	\begin{tabular}{cccccc} 
		\hline
	   Mass &Z &Stage &Lifetime    &$L_\nu$   &$L$ \\
	                & &      &years     &erg s$^{-1}$ &erg s$^{-1}$ \\
		\hline
		150  &0.002 &H-burn  &2.194E+06   &0.652514E+33   &2.30178E+40\\
		     &      &He-burn &2.778E+05   &0.112373E+39   &2.24939E+40\\
		     &      &C-burn  &3.474E+03   &0.344462E+46   &2.46073E+40 \\
		     &      &Ne-burn &8.958E-03   &0.200884E+47   &2.47209E+40\\
	         &      &O-burn  &2.937E-03   &0.728612E+48   &2.51805E+40   \\
		\hline
		200  &0.002 &H-burn  &2.639E+06    &0.293798E+35   &2.87121E+40\\
		     &      &He-burn &2.478E+05    &0.321590E+39   &2.83181E+40\\
		     &      &C-burn  &2.158E+03    &0.163280E+47   &3.06242E+40\\
		     &      &Ne-burn &1.389E-03    &0.281351E+47   &3.06948E+40  \\
	         &      &O-burn  &2.473E-03    &0.108831E+49   &3.12654E+40  \\
	     \hline
		300 &0.002 &H-burn  &2.341+E06     &0.293092E+35   &3.41244E+40 \\
		    &      &He-burn &2.446+E05     &0.453238E+39   &3.35015E+40 \\
		    &      &C-burn  &1.975E+03     &0.237088E+47   &3.58975E+40 \\
		    &      &Ne-burn &1.004E-03     &0.381777E+47   &3.59803E+40   \\
	        &      &O-burn  &2.085E-03     &0.148704E+49   &3.63969E+40   \\
	     \hline
		500 &0.006 &H-burn  &2.112E+06    &0.576674E+34     &1.74207E+40\\
		    &      &He-burn &2.688E+05    &0.179458E+39     &1.52077E+40 \\
		    &      &C-burn  &2.428E+03    &0.147446E+46     &1.70241E+40 \\
		    &      &Ne-burn &1.585E-02    &0.128171E+47     &1.72609E+40   \\
	        &      &O-burn  &1.679E-03    &0.303914E+47     &1.73806E+40  \\
	     \hline
	\end{tabular}

\end{table}

\section*{Acknowledgements}

HA Kassim and N Yusof acknowledge the Fundamental Research 
Grant Scheme grant number FP042-2018A and NS Ahmad also acknowledges the FP042-2018A grant and University of Malaya for the University of Malaya Student Financial Aid for her postgraduate studies.  LG Garba wishes to acknowledge Yusuf Maitama Sule University, Kano, Nigeria. 

\section*{Data Availability}

The data underlying this article will be shared on reasonable request to the corresponding author. 



\bibliographystyle{mnras}
\bibliography{neutrino_vms} 






\bsp	
\label{lastpage}
\end{document}